\documentstyle[12pt]{article}

\begin{document}
\begin{titlepage}
\begin{center}

June 20, 1998      \hfill    LBNL-41813 \\

\vskip .5in

{\large \bf Comment on ``Possibility of quantum mechanics being nonlocal.''}
\footnote{This work was supported by the Director, Office of Energy 
Research, Office of High Energy and Nuclear Physics, Division of High 
Energy Physics of the U.S. Department of Energy under Contract 
DE-AC03-76SF00098.}
\vskip .50in
Henry P. Stapp\\
{\em Lawrence Berkeley National Laboratory\\
      University of California\\
    Berkeley, California 94720}
\end{center}

\vskip .5in

\begin{abstract}

A recent proof, formulated in the symbolic language of modal logic, shows 
that a well-defined formulation of the possibility mentioned in the 
title is answered affirmatively. In the paper being commented upon several 
proposals were made about how to translate this symbolic proof into prose, 
and it was concluded, on the basis of those proposed translations, that 
either the proof was invalid or that an unwarranted reality assumption was 
made. However, those interpretations deviate in small but important ways from
the precise logical path followed in the proof. It is explained here how
by staying on this path one avoids the difficulties that those deviations 
engendered.

\end{abstract}
\medskip
\end{titlepage}

\renewcommand{\thepage}{\roman{page}}
\setcounter{page}{2}
\mbox{ }

\vskip 1in

\begin{center}
{\bf Disclaimer}
\end{center}

\vskip .2in

\begin{scriptsize}
\begin{quotation}

This document was prepared as an account of work sponsored by the United
States Government. While this document is believed to contain correct 
 information, neither the United States Government nor any agency
thereof, nor The Regents of the University of California, nor any of their
employees, makes any warranty, express or implied, or assumes any legal
liability or responsibility for the accuracy, completeness, or usefulness
of any information, apparatus, product, or process disclosed, or represents
that its use would not infringe privately owned rights.  Reference herein
to any specific commercial products process, or service by its trade name,
trademark, manufacturer, or otherwise, does not necessarily constitute or
imply its endorsement, recommendation, or favoring by the United States
Government or any agency thereof, or The Regents of the University of
California.  The views and opinions of authors expressed herein do not
necessarily state or reflect those of the United States Government or any
agency thereof or The Regents of the University of California and shall
not be used for advertising or product endorsement purposes.
\end{quotation}
\end{scriptsize}

\vskip 2in

\begin{center}
\begin{small}
{\it Lawrence Berkeley Laboratory is an equal opportunity employer.}
\end{small}
\end{center}

\newpage
\renewcommand{\thepage}{\arabic{page}}
\setcounter{page}{1}

One of the great lessons of quantum theory is that utmost caution must be 
exercised in reasoning about hypothetical outcomes of unperformed 
experiments. Yet Bohr [1] did not challenge the argument of Einstein,
Podolsky, and Rosen [2] on the grounds that it was based on the 
simultaneous consideration of mutually exclusive possibilities.
Rather he challenged the underlying EPR presumption that an 
experiment performed locally on one system would occur ``without in 
any way disturbing '' a faraway system. Bohr's own ideas rested 
heavily on the idea that experimenters could freely choose between
alternative possible measurements, and the core of his answer to EPR was that 
although {\it ``... there is in a case like that just considered no 
question of a mechanical influence of the system under investigation during 
the last critical stage of the measuring procedure.''}...``there is 
essentially the question of {\it an influence of the very conditions which 
define the possible types of predictions about future behavior of the 
system.''} 

The adequacy of Bohr's answer and the nature of his intermediate 
position on the question of these influences have been much debated.
The issue is of fundamental importance, because it concerns the nature 
of the causal structure of quantum theory, and its compatibility with an 
idea, drawn from the theory of relativity, that no influence of any kind 
can act backward in time in any frame.

The background is this. In relativistic classical physical
theory the actual physical world is conceived to be one of a host of 
possible worlds that all obey the same laws of nature. With fixed initial 
conditions one can, by making a change in the Lagrangian in small 
space-time region, shift from the actual world to a neighboring possible
world, and prove that the effects of this change are confined to times that
lie later than the cause in every Lorentz frame. The change in the 
Lagrangian in the small region can be imagined to alter an experimenter's
choice of which experiment he will soon perform in that region.

An analogous result holds in quantum field theory. However, in the quantum 
case that result is not the whole story: the eventual occurrence of 
the individual outcome must be described. In that connection, Bohr [3]
mentions a discussion at the 1927 Solvay conference as to whether, as Dirac
proposed, we should say that we are `` concerned 
with a choice on the part of `Nature' or, as suggested by Heisenberg, we 
should say that we have to do with a choice on the part of the `observer'  
constructing the measuring instruments and reading their recording.'' 
It is just the possible effect of such a choice, made by an experimenter in one
region, upon the outcome that appears to the observers located in another 
region that is the issue here.

The question, more precisely, is this: Is it possible to maintain  
in quantum mechanics, as one can in classical mechanics, the theoretical idea 
that the one real world that we experience can be imbedded in a set of 
possible worlds, each of which obeys the known laws of physics, if (1), the 
experimenters can be imagined to be able to freely choose between the 
different possible measurements that they might perform, and (2), no such 
free choice can have any effect on anything that lies earlier in time in 
some Lorentz frame. 

It was proved in  [4] that with a sufficently broad definition of 
``anything'' the answer to this question is no.

To obtain rigorous results in this domain it is necessary to formulate
arguments within a formal logic, where each separate statement can be 
stated precisely, and the rules of inference
connecting them are spelled out  exactly.

A framework has been developed by philosophers and logicians for dealing, 
in a logically consistent way, with relationships between the real 
and possible worlds. It is called modal logic. It is designed to
formalize in logically coherent rules what we normally mean by 
statements pertaining to these hypothetical worlds and their connections
to the unique actual world. Although there are several versions of modal 
logic, which differ on fine points [5], they all adhere to certain general
rules.

The proof given in [4] follows the general rules of modal logic. However, 
that does not guarantee that the proof is satisfactory. For modal logic was 
created by philosophers and logicians within a context in which the
actual world and the physical laws that governed it were believed 
to be basically similar to what was imagined to exist in classical physics. 
But the quantum world is profoundly different from this classical 
idealization. Hence the entire question of the appropriate logic must be 
re-examined in a quantum context, where the very idea of the truth of 
statements about hypothetical worlds is greatly curtailed relative to 
classical physics. Utmost care must be taken not to introduce any notion of 
reality that is  contrary to the philosophical principles of quantum 
theory. Thus I shall here avoid any reliance on the symbols and concepts of 
modal logic, and, while actually conforming to the general rules, shall
speak directly to quantum physicists. 

The philosophy of Niels Bohr, as normally understood, allows 
one to imagine that a free choice made by an experimenter about which 
experiment to perform would leave undisturbed an outcome that has actually 
already appeared earlier to the observers of some other experiment.
That notion is one of the ideas that is under scutiny here.

The first locality condition used in the proof expresses this condition.
It is called LOC1. It states that 
if an experiment L2 is actually performed in a spacetime region L, and an 
experiment R2 is actually performed in a faraway region R that lies 
later in time than L (in some frame), and if an outcome c actually 
appears to the observers stationed in the earlier region L, then that 
same result c would appear to the observers in that earlier region 
L also in the alternative {\it possible world} in which everything is left
unchanged except for (1), the free choice made later in time by the 
experimenter in R, and (2), the consequences of that later-in-time change: 
LOC1 asserts that the later free choice in R has no  effect on the outcome 
that has already appeared earlier to the observers located in region L.

The argument in reference [4], stated here in words, rather than the 
symbols of modal logic, begins as follows:

Suppose the actual situation is one in which L2 and R2 are performed and the
outcome g appears to the observers in R. Then a prediction of quantum 
theory, in the Hardy case under consideration in [4], entails that the 
outcome actually appearing to the observers in L must be c.
But according to LOC1 this outcome c actually appearing to the observers in L 
would not be disturbed if the later free choice in R would be to perform R1 
instead of R2: the outcome appearing to the observers in L would still be
c. But then if the laws of quantum theory are assumed to hold not only in 
the single unique world that is actually created by our free choices but 
also in the alternative possible worlds that would be created if our 
alternative free choices had been different, then another prediction of 
quantum theory in this Hardy case entails that the outcome appearing to 
the observers in R, in this alternative possible situation (in which R1 
is performed instead of R2) must be f. 

This result is expressed in line 5 of my proof, which in prose reads:

LINE 5: If L2 is performed then SR is true,

\noindent where SR is the statement:

SR: ``If R2 is performed in R and outcome g appears to the observers in R, 
then if R1, instead of R2, had been performed in R the outcome f would 
have appeared to the observers in R.''

The form of this claim in line 5 is the same as a typical claim in classical
mechanics: if the result of a certain measurement 1 is, say, g, then the 
deterministic laws of physics may allow one to deduce that if some 
alternative possible measurement 2 had been performed, instead of 1, then 
the result of that measurement 2 would necessarily have been f: knowledge 
of what happens in an actual experimental situation may, with the help of
theory, allow one to infer what would have happened if one had performed, 
instead, a different experiment.

Note that no outcome of any actually unperformed measured is asserted to 
exist unless the specific outcome is uniquely fixed by the explicitly 
stated assumptions.

Note also that the assumptions in line 5 pertaining to region L do not 
include the condition that outcome c appears to the observers in L: 
that condition is implied by a prediction of quantum theory and the 
explicitly stated conditions, namely that L2 and R2 are actually performed,
and that outcome g actually appears to the observers in R.

This prose description of this first part of the argument is clear and 
direct, and it conforms to the meanings formalized by the rules modal 
logic.  As mentioned above, these rules may be contaminated by philosophical 
prejudices drawn from the classical conception of the nature of reality. 
However, Niels Bohr, in order to have a secure basis for his own reasoning,
and for the reasoning of practical scientists, insisted that no special 
non-classical-type of logic or reasoning is needed to deal with our 
descriptions at the level of possible experimental set-ups, and our 
observations of their outcomes. No other kind of description enters into 
my argument. 

In any case, I have here laid out the argument in ordinary language,
for physicists to see.

Unruh [6] proposed various different interpretations of some of the 
statements in my proof, and encountered serious difficulties. 
	
Of the interpretations of LOC1 offered by Unruh, the one closest to the one 
occurring in my proof is the one he describes first: ``On face value this 
is just the unexceptional statement that if L2 is measured to have value 
c then the truth of having obtained that value is independent of what 
is (or will be) measured at R''. My statement is only slightly different: 
``This is just the unexceptional statement that if L2 is performed and the 
outcome appearing to the observers in L is c, then this latter fact is 
independent of which measurement will later be performed in R: the later
free choice by the experimenter in R does not disturb what has actually
appeared earlier to the observers in region L.'' 

Unruh claims that ``this meaning of LOC1 is insufficient to derive his 
[Stapp's] conclusion, since it demands that L2 had actually been measured 
and had the given outcome.''

To understand this objection we must turn to the second locality condition,
LOC2, and its application, for that is where the condition L2 is relaxed.

The application of LOC2 is connected to line 5, which says that if L2 is 
performed then statement SR is true..

Suppose the experiment L2 is performed in a spacetime region that lies much
later in time (in some frame) than all points in the region R, in which all 
of the possible events referred to by SR lie. And suppose the later choice 
between L1 and L2 is really free: i.e., that this choice is independent of 
everything earlier. And suppose the assertion that ``L1 is 
performed instead of L2'' means specifically that nothing is changed relative
to the real situation in which L2 is performed except for L2 and the 
consequences of the change of the free choice in L from L2 to L1. 

Then the demand that there be no backward-in-time influence of any kind
requires that SR cannot be true in the actual world in which L2 is performed,
but be untrue if L1 is performed instead of L2: such a difference would
constitute {\it some sort} of backward-in-time influence.

LOC2 is, accordingly, the assertion that if SR is true under condition L2,
then it is also true if the free choice made later in  L is to perform L1 
instead of L2: the change in the later free choice in L cannot, without acting
backward in time, disturb the truth of a statement whose truth or falsity is 
determined by a relationship between possible events that are all located 
earlier in time.

We may now return to Unruh's claim that ``this meaning of LOC1 is 
insufficient to derive his [Stapp's] conclusion, since it demands that 
L2 had actually been measured and had the given outcome.''

My proof is based squarely on the premise that L2 is actually performed.
The other conditions are that R2 is actually performed and that the outcome g 
actually appears to the observers stationed in R. These three conditions 
entail, by virtue of a prediction of quantum theory---accepted as valid in 
the actual world---that the outcome c actually appears to the observers 
stationed in L. This entailment is a consequence of the von Neumann type 
argument that Unruh has given as an example of reasoning that is valid in 
the quantum context. So my proof satisfies exactly the conditions that 
Unruh demands, namely that L2 is actually measured and has the actual 
outcome c. 

It is important, in following a logical argument, to proceed 
step-by-step in a logical progression that leads from the assumptions to 
the conclusions. In my proof the assumption LOC1 is used to get line 5. 
In that application of LOC1 the experiment performed in region L is fixed 
to be the actually performed experiment L2. And the premise of SR, 
together with a prediction of quantum theory, allows one to conclude 
that the outcome actually observed by the observers in L is the outcome c. 
So all the conditions for the applicability of LOC1 are satisfied. 
LOC1 is not used thereafter. 

The answer, therefore, to Unruh's claim about the insufficiency of LOC1
is that in my proof the assumption LOC1 is used only under conditions where 
its use is justified, and in particular only under the condition that L2 
is performed in L. The effects of the switch from the real L2 to the 
hypothetical L1 are restricted only by the second locality condition, LOC2.

Unruh formulates also another objection to LOC1. He suggests that, instead 
of the first interpretation of LOC1 that he proposed, which is the one most
similar to the one used in my proof, that perhaps LOC1 means that ``if 
one is somehow able to infer that L2 has value c then it remains true that 
L2 has value c under replacement of R2 with R1 even if the outcome of the 
measurement R2 was crucial in drawing the inference that L2 has value c.'' 
He says that: ``This interpretation of LOC1 is. I would argue, a form of 
realism, in that it claims that the value to be ascribed to L2 is 
independent of the evidence used to determine that value.''

This argument is based essentially on a reversal, relative to my proof, 
of what is real and what is hypothetical.

The evidence used to ascribe the value c to L2 is the actual sensory 
evidence of the appearance of outcome c to the observers stationed in 
region L. Thus the evidence that the outcome c appears to observers in L---
which is my way of expressing Unruh's condition on the value---is precisely
that the outcome c really appears to the observers in L. Hence the 
relationship between the {\it evidence} and the {\it value} is the 
relationship of {\it identity}, not {\it independence}. It is only 
under this condition---namely that this sensory evidence really 
exists---that the locality condition LOC1 specifies that 
what really appears to the observers in the earlier region L would not 
be disturbed by a shift to a theoretically possible (but unreal) world 
in which everything is exactly the same as in the actual world except for 
{\it consequences} of the change in the later free choice to perform 
R1 instead  of R2. The locality condition LOC1 that we are testing here 
is precisely that this later switch from R2 to R1, which certainly
would produce an imaginary world in which no outcome of R2 would appear to 
anyone, would nevertheless leave undisturbed what has, in the real world, 
already happened in the earlier region L.

This interchange of real and imaginary occurs repeatedly into Unruh's 
argument. For example, between his equations (5) and (6)  he says: 
``In particular, the truth of the statement made about system A which 
relies on measurement made on system B and the correlations which have 
been established between A and B in the state of the joint system is 
entirely dependent on the truth of the actual measurement which has been 
made on system B. To divorce them is to effectively claim that the 
statement made about A can have a value in and of itself, and independent 
of measurements which have been made on A. This notion is equivalent to 
asserting the reality of the statement about A independent of measurements, 
a position contradicted by quantum mechanics.''

There appears to be some nonstandard usages of the words ``truth'' and
``reality''. But in any case a key feature of my proof is that the 
measurement on system B [i.e. R2] is really performed, and that all of the 
consequences of this real action really occur. This feature is 
deeply ingrained in the modal-logic formalism, and is an important part 
of what makes that logic---and, by extension, my argument---cohesive and 
coherent. Deviations from this  reality structure upsets the logical 
structure of my argument.

Unruh's objections described above pertain to LOC1. But he raises an 
objection also to LOC2.

He says: ``If it were true that one could deduce solely from the fact that 
a measurement had been made at L that some relation on the right must hold,
then I would agree that this requirement [LOC2] would be reasonable.''

The other assumptions, including LOC1, are of course needed. But, given 
those other assumptions, which I have specified, line 5 [L2 implies SR] 
asserts that SR can be ``deduced solely from the fact that'' L2 is 
performed: one does not need to {\it assume} that the outcome appearing 
to the observers stationed in L is c. 

Unruh says that he ``would agree that this requirement would be reasonable''
if the stated premise were true, i.e.,  ``If it were true that...[L2 implies 
SR]''.  As discussed above, he had previously given arguments that led him to 
believe that premise to be false. But the analysis just concluded shows, I 
believe, that those  criticisms were linked to deviations from the path
followed in my proof. Given the validity of line 5 [L2 implies SR], 
Unruh's statement acknowledges that LOC2 is all right.

Immediately after this qualified endorsement of LOC2  Unruh says: 
``However, if the truth of the relation on the right hand side 
depended not only on which measurement had been made 
[I would say ``will be made''] on the left, but also on the actual value 
obtained on the left, then no such locality condition would obtain.'' 

It is certainly  true that if the truth of SR depended upon which 
measurement is made on the left, then that fact alone, by itself, simply 
by definition, is enough to make LOC2 false; for LOC2 asserts that, on the
contrary, the truth of SR is {\it independent} of which measurement is 
performed in L. Hence this second statement is true by definition.

Unruh then goes on to say: ``If it is the value [c] obtained on the left...
which allows one to deduce [the truth of] the relation [SR] on the right,
then [the truth of] that relation [SR] on the right cannot be independent
of what is measured on the left, but rather is tied to that measured value. 
To assume otherwise, to assume that the [truth of the] relations between 
possible measurements on the right are independent of the values on the left
that were used to derive [the truth of] those relations, is, in my opinion, 
simply another form of realism.'' [I have inserted the contents of the 
square brackets to make more precise what I believe Unruh to be saying.]
 
There is an ambiguity in the meaning of `depend upon' in Unruh's assertion 
that the truth of SR ``cannot be independent of... .'' What the truth of SR 
depends upon might mean the basic conditions that define whether SR is true. 
Or it might mean some particular condition that is sufficient to ensure that 
these basic conditions are satisfied. Or it might mean some third condition 
that enters into a proof that these basic conditions are satisfied under that 
particular condition. 

Unruh's statement uses to the third meaning of ``depends upon'', whereas
the meaning that is rationally needed in my proof is the first meaning.

A key function of logic is to organize the 
reasoning process so it does not have to carry along the entire proof of 
a statement in order to give that statement a well defined meaning. Indeed,
one generally sets out to prove the truth of some statement without even 
knowing whether there is a proof. Thus the definition of the conditions
under which a statement is true needs to be separable from a proof that the 
statement is true.

The {\it proof} of line 5 certainly depends on the fact that, under the
actual conditions specified in line 5, the outcome c actually occurs.
There is no doubt about that. But this proof that SR holds under condition L2 
completely collapses in the context where L1 is performed instead of L2: if 
the proof did not collapse then LOC2 would add nothing.  So all evidence 
for the truth of SR coming from the {\it proof} of line 5 vanishes if L1 
is performed instead of L2. The next step in my argument is based on a 
completely different consideration. It is severed from earlier part by the 
fact that the ``meaning'' of line 5 is separate from the ``proof'' of
line 5. The meaning of line 5 is fixed prior to its proof: to confound the 
proof of a statement with its meaning [in the sense of the defining
conditions for the statement to be true] is to abandon rational 
thinking itself. 

In view of the {\it meaning} of statement 5, it is reasonable to 
assert, as a direct formulation of the idea that free choices can have no 
effects backward in time, the following demand:  if it is known [e.g., by 
means of some valid argument] that a certain relationship among elements 
confined to an earlier region R must hold provided the later free choice in L 
is L2, then {\it that same relationship} must continue hold if the later free 
choice were to go the other way, provided there were no other changes except
for consequences of the change in this later free choice. This demand is LOC2,
and the pertinent {\it relationship} that should continue to hold is the 
{\it defining condition} for SR to be true.

In the end, the significance of the proof lies in how it can be used. The 
purpose of this proof is to place a stringent condition on the possibilities 
of modeling quantum theory. Line 5 says [under the conditions that the choices
made by experimenters can be treated, in this context, as free choices, and
that a later-in-time change of such a choice (e.g., in R) cannot produce, as 
a consequence,  a change in an outcome that has already appeared to the actual
observers (e.g., in L)] that if L2 is performed earlier, then a certain 
specified relationship must hold between the outcomes of two alternative 
possible measurements in R. LOC2 then adds the independent locality condition 
that the existence of a constant relationship between outcomes of possible 
experiments in R (now regarded as earlier) cannot depend on which experiment 
is freely chosen later: i.e., on whether L2 or L1 is chosen in region L. The 
logical contradiction that ensues appears to rule out any model that 
reproduces the predictions of quantum theory but that forbids the free choices
made by experimenters from having {\it any sort of effects} that act backward 
in time in some frame. Thus the proof is not something that simply can be  
cast aside by some verbal convention: it places a severe condition on any
model of nature that produces observations that agree with the predictions of 
quantum theory, and in which the choices made by experimenters can be treated 
as free variables.


\begin{thebibliography}{99}
\bibitem{kn:bohr}  N. Bohr, Phys. Rev. {\bf 48}, 696 (1935).
\bibitem{kn:epr}  A. Einstein, B. Podolsky, and N. Rosen, Phys. Rev. {\bf 47},
777 (1935).  
\bibitem{kn:bi} N. Bohr in Albert Einstein: Philosopher-Scientist.
P.A. Schilpp, Tudor, New York, 1951. pg.223.
\bibitem{kn:hps} H.P. Stapp, Am. J. Phys. {\bf 65}, 300 (1997).
\bibitem{kn:dl}  David Lewis, Philosophical Papers, Vol II, Oxford U.P.
(1986); Counterfactuals, Blackwell Press, Oxford, (1973).
\bibitem{kn:unruh} W. Unruh, Phys. Rev. A, to appear. (quant-ph/9710032).
\end{thebibliography}
\end{document}